\documentclass[preprint, superscriptaddress,showpacs,preprintnumbers,amsmath,amssymb,prb]{revtex4}
\usepackage{graphicx}
\usepackage{amsmath}

\begin{document}

\thispagestyle{empty}

\title{Conductivity of pure graphene: Theoretical approach using the polarization tensor}

\author{
G.~L.~Klimchitskaya}
\affiliation{Central Astronomical Observatory at Pulkovo of the Russian Academy of Sciences,
Saint Petersburg,
196140, Russia}
\affiliation{Institute of Physics, Nanotechnology and
Telecommunications, Peter the Great Saint Petersburg
Polytechnic University, St.Petersburg, 195251, Russia}

\author{
V.~M.~Mostepanenko}
\affiliation{Central Astronomical Observatory at Pulkovo of the Russian Academy of Sciences,
Saint Petersburg,
196140, Russia}
\affiliation{Institute of Physics, Nanotechnology and
Telecommunications, Peter the Great Saint Petersburg
Polytechnic University, St.Petersburg, 195251, Russia}

\begin{abstract}
We obtain analytic expressions for the conductivity of pristine (pure)
graphene in the framework of the Dirac model using the polarization
tensor in (2+1)-dimensions defined along the real frequency axis. It
is found that at both zero and nonzero temperature $T$ the in-plane and
out-of-plane conductivities of graphene are equal to each other with
a high precision and essentially do not depend on the wave vector.
At $T=0$ the conductivity of graphene is real and equal to
$\sigma_0=e^2/(4\hbar)$ up to small nonlocal corrections in accordance
with many authors. At some fixed $T\neq 0$ the real part of the
conductivity varies between zero at low frequencies $\omega$ and $\sigma_0$
for optical $\omega$. If $\omega$ is fixed, the conductivity varies between
$\sigma_0$ at low $T$ and zero at high $T$. The imaginary part of the
conductivity of graphene is shown to depend on the ratio of $\omega$ to $T$.
In accordance to the obtained asymptotic expressions, at fixed $T$ it varies
from infinity at $\omega=0$ to a negative minimum value reached at some
$\omega$, and then approaches to zero with further increase of $\omega$.
At fixed $\omega$ the imaginary part of the conductivity varies from zero
at $T=0$, reaches a negative minimum at some $T$ and then goes to infinity
together with $T$. The numerical computations of both the real and
imaginary parts of the conductivity are performed. The above results are
obtained in the framework of quantum electrodynamics at nonzero temperature
and can be generalized for graphene samples with nonzero mass gap parameter
and chemical potential.
\end{abstract}
\pacs{72.80.Vp, 73.63.-b, 65.80.Ck}

\maketitle

\section{Introduction}

It is generally recognized that graphene and other two-dimensional materials
have opened up new fields in both fundamental and applied research.\cite{1}
Unlike ordinary three-dimensional materials, quasiparticles in pure graphene
are massless charged fermions. At energies below a few eV they possess the
linear dispersion relation and obey the Dirac equation, where the speed of light
$c$ is replaced with the Fermi velocity $v_F\approx c/300$. This makes possible
to use graphene for testing the Klein paradox,\cite{2} the effect of particle
creation from vacuum by electric field,\cite{3,4,5} the Casimir effect\cite{6,7}
and other phenomena of relativistic quantum field theory.

Numerous applications of graphene in condensed matter physics and nanotechnology
are based on its unusual electronic properties.\cite{8} Among them the electrical
conductivity of graphene has received the most study both theoretically and
experimentally.\cite{9,10,11} The striking result from the study of conductivity
of Dirac fermions\cite{12} is the existence of universal and frequency independent
conductivity of graphene
\begin{equation}
\sigma_0=\frac{\pi e^2}{2 h}=\frac{e^2}{4\hbar}
\label{eq0}
\end{equation}
\noindent
in the limiting case of vanishing temperature and disorder. It was noted also\cite{12}
that if the frequency vanishes first and disorder second, the obtained minimal dc
conductivity is different from $\sigma_0$. In this case for graphene one obtains
$\sigma_{\rm dc}=4e^2/(\pi h)=8\sigma_0/\pi^2$. The existence of nonzero universal
conductivity $\sigma_0$ is usually considered as surprising because in the limit
of zero temperature there are no charge carriers, no scattering and no dissipation
processes. Theoretically the conductivity of graphene was investigated in many
papers using the current-current correlation functions, the Kubo formalism and
Boltzmann's transport theory.\cite{13,14,15,16,17,18,19,20,21,22,23,24,25,26,27,28}
It was noted\cite{11,21} that the values of minimal
conductivity obtained by different
authors vary depending on the order of limiting transitions used in different
theoretical approaches. At the same time, measurements of the conductivity of
graphene\cite{29,30,31,32,33} result in somewhat larger values than the theoretical
predictions.\cite{11,22}

During the last few years much attention was attracted also to investigation of the
Casimir force between two graphene sheets, a graphene sheet and a plate made of
ordinary material and between graphene-coated substrates. In so doing, the
density-density correlation functions of graphene in the random phase approximation,
the spatially nonlocal dielectric permittivities and other calculation methods have
been used.\cite{6,34,35,36,37,38} It was shown, however, that the most complete
results for the Casimir force in graphene systems are obtained using the polarization
tensor of graphene in (2+1)-dimensional space-time.\cite{39,40} Using this tensor
defined at pure imaginary Matsubara frequencies, the Casimir effect in many graphene
systems has been investigated\cite{41,42,43,44,45,46} and an equivalence to some other
calculation methods has been proved.\cite{47}

It should be taken into account that thermal quantum field theory provides the
rigorous and straightforward formalism allowing calculation of the polarization
tensor for different systems. Because of this, it is desirable to investigate
as many physical phenomena as possible using this approach. Keeping in mind that
calculation of the Casimir force in the framework of the Lifshitz theory\cite{48}
is based on the reflection coefficients, it was interesting to study the reflectivity
properties of graphene using the polarization tensor. For this purpose it was
necessary to obtain an alternative representation for the polarization tensor of
graphene valid not only at the pure imaginary Matsubara frequencies (as the one of
Ref.~[\onlinecite{40}]), but along the real frequency axis as well. This problem was
solved in Ref.~[\onlinecite{49}], and the reflectivity properties of pure
graphene were studied over the wide range of frequencies. The obtained representation
for the polarization tensor was also used to investigate the reflectivity properties
of gapped graphene, graphene-coated substrates,\cite{50}
and large thermal effect in the Casimir force
between two graphene sheets.\cite{51,52}

One further physical quantity of great interest directly expressed via the polarization
tensor is the electrical conductivity.
There are lots of papers in the literature where the conductivity of graphene is studied
using different theoretical approaches (some of them are cited above). However, even for
the case of pure graphene, where the effects of disorder are not taken into account,
different values of the conductivity were obtained.
Keeping in mind that the most of previous literature on the subject used various
phenomenological models,
 it is reasonable to investigate the conductivity of pure graphene again
starting from the first principles of thermal quantum field theory.
In this paper we calculate both the longitudinal (in-plane) and transverse (out-of-plane)
conductivities of pure (pristine) graphene using the formalism of the polarization tensor
at arbitrary temperature. We obtain general analytic expressions and approximate
asymptotic results for both the real and imaginary parts of the conductivity of graphene.
The analytic results are accompanied by numerical computations over the wide regions of
frequency and temperature. It is shown that at zero temperature the conductivity of
graphene is real and takes the universal value $\sigma_0$ up to small nonlocal
corrections. The real part of the conductivity of graphene taken at fixed temperature
goes to zero and to $\sigma_0$ with decreasing and increasing frequency, respectively.
We also show that the real part of the graphene conductivity at fixed frequency goes to
$\sigma_0$ when the temperature vanishes and to zero with increasing temperature.
At any nonzero temperature the conductivity of graphene has both the real and imaginary parts.
We demonstrate that at some relationship between frequency and temperature the imaginary
part of the conductivity of graphene drops to zero. At fixed temperature, the imaginary
part of the conductivity goes to infinity with vanishing frequency and to zero with
increasing frequency. At fixed frequency, the imaginary
part of the graphene conductivity goes to zero with vanishing temperature and to infinity
with increasing temperature. We also compare the obtained results with those found in the
literature using some other theoretical approaches.

The paper is organized as follows. In Sec.~II the general formalism connecting the
conductivity of graphene with the polarization tensor is presented. In Sec.~III the real
part of the conductivity of graphene is investigated at both zero and nonzero temperature.
Section IV is devoted to the imaginary part of the graphene conductivity.
It presents both the asymptotic expressions and numerical results. In Sec.~V the reader
will find our conclusions and discussion.

Throughout the paper we preserve the fundamental constants in all formulas, but measure
frequencies in the units of energy.

\section{General formalism}

The polarization tensor of graphene at temperature $T$ in the one-loop approximation
is defined as\cite{40,49}
\begin{eqnarray}
&&
\Pi_{\mu\nu}(\omega,\mbox{\boldmath$k$},T)=-8\pi e^2\frac{k_BT}{\hbar c^2}
\label{eq1} \\
&&~~~~~~~~~~\times
\sum_{n=-\infty}^{\infty}\int\frac{d\mbox{\boldmath$q$}}{(2\pi)^2}\,
{\rm tr}S(q_n)\tilde{\gamma}_{\mu}S(q_n-k)\tilde{\gamma}_{\nu},
\nonumber
\end{eqnarray}
\noindent
where $k_B$ is the Boltzmann constant, $e$ is the electron  charge, the trace is
taken over the $\gamma$ matrices, and the spinor propagator is given by
\begin{equation}
S(q)=\frac{1}{i\tilde{\gamma}_{\alpha}q^{\alpha}-mc/\hbar}.
\label{eq2}
\end{equation}
\noindent
In Eq.~(\ref{eq1}) $k\equiv k^{\alpha}=(\omega/c,\mbox{\boldmath$k$})$ is the
3-dimensional wave vector of an external photon, where
$\mbox{\boldmath$k$}=(k^1,k^2)$ are the components of the wave vector in the plane
of graphene, and $q\equiv q^{\alpha}=(q^0,\mbox{\boldmath$q$})$ is the
3-dimensional wave vector of a loop electronic excitation.
In the framework of the thermal quantum field theory in the Matsubara formulation
the continuous zeroth component of the electronic wave vector should be replaced
with the discrete half-integer pure imaginary quantity
\begin{equation}
q_n^0=2\pi i\frac{k_BT}{c\hbar}\left(n+\frac{1}{2}\right), \quad
q_n\equiv q_n^{\alpha}=(q_n^0,\mbox{\boldmath$q$}).
\label{eq3}
\end{equation}
\noindent
The $\tilde{\gamma}$ matrices in Eqs.~(\ref{eq1}) and (\ref{eq2}) are connected with
the standard $4\times 4$ Dirac $\gamma$-matrices by the relations
\begin{equation}
\tilde{\gamma}_{\alpha}=\eta_{\alpha}^{\,\beta}\gamma_{\beta}, \quad
\eta_{\alpha}^{\,\beta}\equiv{\rm diag}(1,\tilde{v}_F,\tilde{v}_F),
\label{eq4}
\end{equation}
\noindent
where $\tilde{v}_F=v_F/c$ and all the Greek indices here and above take the values
0,\,1,\,2.

It is important that all components of the polarization tensor of graphene can be
expressed\cite{40} via $\Pi_{00}$ and $\Pi_{\rm tr}\equiv\Pi_{\mu}^{\,\mu}$.
Below it is more convenient to use the following combination:
\begin{equation}
\Pi(\omega,\mbox{\boldmath$k$},T)\equiv k^2\Pi_{\rm tr}(\omega,\mbox{\boldmath$k$},T)
+\left(\frac{\omega^2}{c^2}-k^2\right)\Pi_{00}(\omega,\mbox{\boldmath$k$},T)
\label{eq5}
\end{equation}
\noindent
instead of $\Pi_{\rm tr}$, where, starting from here,
$k\equiv |\mbox{\boldmath$k$}|=\sqrt{k_1^2+k_2^2}$, i.e., is the magnitude of the
projection of the photon wave vector on the plane of graphene. It was shown\cite{40,49}
that the components of the polarization tensor depend only on $\omega$,
$k= |\mbox{\boldmath$k$}|$ and $T$.

There are the familiar expressions connecting the
in-plane ($\sigma_{||}$) and out-of-plane ($\sigma_{\!\bot}$) conductivities of
the two-dimensional graphene sheet with its dynamical polarizabilities and dielectric
permittivities \cite{37,52a}
\begin{equation}
\alpha_{\|(\bot)}(\omega,k,T)\equiv \varepsilon_{\|(\bot)}(\omega,k,T)-1=
\frac{2\pi i\sigma_{\|(\bot)}(\omega,k,T)k}{\omega}.
\label{eq6a}
\end{equation}
\noindent
Using the Maxwell equations and the standard electrodynamic boundary conditions,
one can express  \cite{19,37,52b} the amplitude reflection coefficients on the
surface of graphene via $\alpha_{\|(\bot)}(\omega,k,T)$ or
$\sigma_{\|(\bot)}(\omega,k,T)$ for the two independent polarizations of the electromagnetic
field, transverse magnetic and transverse electric (i.e., for $p$ and $s$ polarizations).
The Maxwell equations allow one to express the same reflection coefficients in terms of
the polarization tensor.\cite{39,40,41}
By comparing both exact expressions for the TM and TE reflection coefficients, the
conductivity of graphene was expressed via the components of polarization tensor.\cite{47}
Using the representation for this tensor
defined over the entire plane of complex
frequencies,\cite{49} rather than that valid at the Matsubara frequencies alone,\cite{40}
the conductivity of graphene takes the form
\begin{eqnarray}
&&
\sigma_{\|}(\omega,k,T)=-i\frac{\omega}{4\pi\hbar k^2}\,\Pi_{00}(\omega,k,T),
\nonumber \\
&&
\sigma_{\bot}(\omega,k,T)=i\frac{c^2}{4\pi\hbar k^2\omega}\,\Pi(\omega,k,T).
\label{eq6}
\end{eqnarray}

It is convenient to represent the polarization tensor as a sum of the zero-temperature
contribution and the thermal correction to it
\begin{eqnarray}
&&
\Pi_{00}(\omega,k,T)=\Pi_{00}^{(0)}(\omega,k)+\Delta_T\Pi_{00}(\omega,k,T),
\nonumber \\
&&
\Pi(\omega,k,T)=\Pi^{(0)}(\omega,k)+\Delta_T\Pi(\omega,k,T).
\label{eq7}
\end{eqnarray}
\noindent
The zero-temperature part can be written in the form\cite{39,49}
\begin{eqnarray}
&&
\Pi_{00}^{(0)}(\omega,k)=ie^2\pi\frac{k^2}{\omega\eta(\omega,k)},
\nonumber \\
&&
\Pi^{(0)}(\omega,k)=-ie^2\pi\frac{\omega}{c^2}k^2\eta(\omega,k),
\label{eq8}
\end{eqnarray}
\noindent
where
\begin{eqnarray}
&&
\eta\equiv\eta(\omega,k)=\sqrt{1-\kappa^2(\omega,k)},
\nonumber \\
&&
\kappa\equiv\kappa(\omega,k)=\tilde{v}_F\frac{ck}{\omega}.
\label{eq9}
\end{eqnarray}
\noindent
Note that the zero-temperature contribution (\ref{eq8})
is the same for both representations of the polarization tensor.\cite{40,49}
For real photons on a mass-shell $k\leq\omega/c$ and, thus,
$\kappa\leq\tilde{v}_F\ll 1$.

The real part of the thermal correction to the polarization tensor\cite{49} can be
most conveniently represented as\cite{50}
\begin{eqnarray}
&&
{\rm Re}\Delta_T\Pi_{00}(\omega,k,T)=\frac{8e^2\omega}{\tilde{v}_F^2c^2}
\sum_{j=1}^{3}Z_{00}^{(j)}(\omega,k,T),
\nonumber \\
&&
{\rm Re}\Delta_T\Pi(\omega,k,T)=\frac{8e^2\omega^3}{\tilde{v}_F^2c^4}
\sum_{j=1}^{3}Z^{(j)}(\omega,k,T).
\label{eq10}
\end{eqnarray}
\noindent
Here, the three dimensionless quantities $Z_{00}^{(j)}$ are defined by
\begin{eqnarray}
&&
Z_{00}^{(1)}(\omega,k,T)\equiv\int_{0}^{1-\kappa}\frac{dy}{e^{\beta y}+1}
\left\{1-\frac{1}{2\eta}\left[\sqrt{(y+1)^2-\kappa^2}+\sqrt{(y-1)^2-\kappa^2}
\right]\right\},
\nonumber \\
&&
Z_{00}^{(2)}(\omega,k,T)\equiv\int_{1-\kappa}^{1+\kappa}\frac{dy}{e^{\beta y}+1}
\left[1-\frac{1}{2\eta}\sqrt{(y+1)^2-\kappa^2}\right],
\label{eq11}\\
&&
Z_{00}^{(3)}(\omega,k,T)\equiv\int_{1+\kappa}^{\infty}\frac{dy}{e^{\beta y}+1}
\left\{1-\frac{1}{2\eta}\left[\sqrt{(y+1)^2-\kappa^2}-\sqrt{(y-1)^2-\kappa^2}
\right]\right\},
\nonumber
\end{eqnarray}
\noindent
where
\begin{equation}
\beta\equiv\beta(\omega,T)=\frac{\omega}{2\omega_T}\equiv\frac{\hbar\omega}{2k_BT}.
\label{eq12}
\end{equation}
\noindent
The quantity $\omega_T=k_BT/\hbar$ has the physical meaning of the thermal frequency.
The three dimensionless quantities $Z^{(j)}$ in Eq.~(\ref{eq10}) are obtained from
the respective quantities $Z_{00}^{(j)}$ in Eq.~(\ref{eq11}) by the following
substitution under the integrals:
\begin{equation}
\sqrt{(y\pm1)^2-\kappa^2}\to\frac{(y\pm 1)^2}{\sqrt{(y\pm 1)^2-\kappa^2}}.
\label{eq13}
\end{equation}

The imaginary part of the thermal correction to the polarization tensor\cite{49}
can be rewritten in the form\cite{50}
\begin{eqnarray}
&&
{\rm Im}\Delta_T\Pi_{00}(\omega,k,T)=\frac{4e^2\omega}{\tilde{v}_F^2c^2\eta}
Y_{00}(\omega,k,T),
\nonumber \\
&&
{\rm Im}\Delta_T\Pi(\omega,k,T)=-\frac{4e^2\omega^3\eta}{\tilde{v}_F^2c^4}
Y(\omega,k,T),
\label{eq14}
\end{eqnarray}
\noindent
where the dimensionless integrals are given by
\begin{eqnarray}
&&
Y_{00}(\omega,k,T)=-\int_{1-\kappa}^{1+\kappa}\frac{dy}{e^{\beta y}+1}
\sqrt{\kappa^2-(1-y)^2},
\nonumber \\
&&
Y(\omega,k,T)=-\int_{1-\kappa}^{1+\kappa}\frac{dy}{e^{\beta y}+1}
\frac{(1-y)^2}{\sqrt{\kappa^2-(1-y)^2}}.
\label{eq15}
\end{eqnarray}

Here we present also an important property of the nonlocal dielectric
permittivities\cite{53}
\begin{equation}
\frac{\omega^2}{c^2}\lim_{k\to 0}
\frac{\varepsilon_{\bot}(\omega,k,T)-\varepsilon_{\|}(\omega,k,T)}{k^2}
=1-\frac{1}{\mu(\omega,T)},
\label{eq17}
\end{equation}
\noindent
where $\mu(\omega,T)$ is the magnetic permeability of a medium. By putting in
our case $\mu=1$ and using Eq.~(\ref{eq6a}), one obtains
\begin{equation}
\lim_{k\to 0}
\frac{\alpha_{\bot}(\omega,k,T)-\alpha_{\|}(\omega,k,T)}{k^2}
=0.
\label{eq18}
\end{equation}
\noindent
Below we use this equality, which means that for the normal incidence the response
functions to $p$ and $s$ polarized light must be equal, to check a consistency of
the obtained analytic
expressions.

\section{Real part of conductivity}

In this section the real part of the conductivity of pure graphene is investigated
with the help of Eqs.~(\ref{eq6})--(\ref{eq9}), (\ref{eq14}) and (\ref{eq15})
expressing it via the polarization tensor. We start from the case of zero
temperature and consider next the properties of the real part of conductivity
at any temperature.

\subsection{Conductivity  via the polarization tensor at zero temperature}

As is seen from Eq.~(\ref{eq8}), the polarization tensor at $T=0$ is pure imaginary.
Substituting Eq.~(\ref{eq8}) in Eq.~(\ref{eq6}), one arrives to the longitudinal
and transverse real conductivities of graphene at zero temperature
\begin{eqnarray}
&&
\sigma_{\|}(\omega,k,0)=
\frac{e^2}{4\hbar\eta(\omega,k)}=
\frac{\sigma_0}{\sqrt{1-\kappa^2(\omega,k)}},
\nonumber \\
&&
\sigma_{\bot}(\omega,k,0)=
\frac{e^2}{4\hbar}\eta(\omega,k)=
{\sigma_0}{\sqrt{1-\kappa^2(\omega,k)}},
\label{eq19}
\end{eqnarray}
\noindent
where $\sigma_0$ is expressed in Eq.~(\ref{eq0}).
The same expressions for $\sigma_{\|(\bot)}$ are presented in Ref.~[\onlinecite{53a}] which contains
a summary of the obtained results at $T=0$ and nonzero chemical potential.
Taking into account Eq.~(\ref{eq9})
and expanding in powers of a small parameter $\kappa$, one arrives at
\begin{eqnarray}
&&
\sigma_{\|}(\omega,k,0)\approx
{\sigma_0}\left[1+\frac{1}{2}\tilde{v}_F^2\left(\frac{ck}{\omega}\right)^2
+O(\tilde{v}_F^4)\right],
\nonumber \\[-2mm]
&&\label{eq20}\\[-2mm]
&&
\sigma_{\bot}(\omega,k,0)\approx
{\sigma_0}\left[1-\frac{1}{2}\tilde{v}_F^2\left(\frac{ck}{\omega}\right)^2
+O(\tilde{v}_F^4)\right].
\nonumber
\end{eqnarray}
\noindent
As is seen from Eq.~(\ref{eq20}), at $T=0$ both conductivities of graphene are
equal to a high precision to the universal constant conductivity $\sigma_0$.
The same value for the minimal conductivity of graphene was obtained by different
authors using the Kubo formula.\cite{12,18,19,21,28,53a,54} The nonlocal corrections to
this result depending on $\omega$ and $k$ are of the order of $10^{-5}$ and can be
neglected.

It is easy to check the fulfilment of Eq.~(\ref{eq18}) at $T=0$. For this purpose we
substitute Eq.~(\ref{eq8}) in Eq.~(\ref{eq6a}) and obtain
\begin{equation}
\alpha_{\|}(\omega,k,0)=
i\frac{e^2\pi}{2\hbar }\frac{k}{\omega\sqrt{1-\kappa^2}},
\quad
\alpha_{\bot}(\omega,k,0)=
i\frac{e^2\pi}{2\hbar }\frac{k}{\omega}\sqrt{1-\kappa^2}.
\label{eq21}
\end{equation}
\noindent
{}From these equations we get
\begin{equation}
\alpha_{\bot}(\omega,k,0)-\alpha_{\|}(\omega,k,0)=
-i\frac{e^2\pi c^2}{2\hbar }\frac{k^3}{\omega^3}
\frac{\tilde{v}_F^2}{\sqrt{1-\kappa^2}}.
\label{eq22}
\end{equation}
\noindent
Taking into account that $\kappa\to 0$ when $k\to 0$, we find that Eq.~(\ref{eq18})
is satisfied.

\subsection{Real part of conductivity via the polarization tensor at nonzero temperature}

As shown in Sec.~II, at nonzero temperature the polarization tensor has both the real and
imaginary parts. The real part of the conductivity under consideration in this section is
determined by the imaginary of the polarization tensor presented in Eq.~(\ref{eq14}).
We start with analytic expressions for the real part of the conductivity of graphene and
then continue with the results of numerical computations.

\subsubsection{Analytic expressions for the real part of conductivity}

According to Eqs.~(\ref{eq6})--(\ref{eq8}) the real parts of the longitudinal and transverse
conductivities of graphene at nonzero temperature are given by
\begin{eqnarray}
&&
{\rm Re}\sigma_{\|}(\omega,k,T)=-i\frac{\omega}{4\pi\hbar k^2}\left[
\Pi_{00}^{(0)}(\omega,k)+i{\rm Im}\Delta_T\Pi_{00}(\omega,k,T)
\right],
\nonumber \\[-2mm]
&&\label{eq23}\\[-2mm]
&&
{\rm Re}\sigma_{\bot}(\omega,k,T)=i\frac{c^2}{4\pi\hbar k^2\omega}\left[
\Pi^{(0)}(\omega,k)+i{\rm Im}\Delta_T\Pi(\omega,k,T)
\right].
\nonumber
\end{eqnarray}
\noindent
Substituting  Eqs.~(\ref{eq8}) and (\ref{eq14}) in  Eq.~(\ref{eq23}), one obtains
\begin{eqnarray}
&&
{\rm Re}\sigma_{\|}(\omega,k,T)=
\frac{\sigma_0}{\eta}\left[1+\frac{4}{\pi\kappa^2}Y_{00}\right],
\nonumber \\[-2mm]
&&\label{eq24}\\[-2mm]
&&
{\rm Re}\sigma_{\bot}(\omega,k,T)=
{\sigma_0}{\eta}\left[1+\frac{4}{\pi\kappa^2}Y\right],
\nonumber
\end{eqnarray}
\noindent
where the integrals $Y_{00}$ and $Y$ are defined in Eq.~(\ref{eq15}).

It is convenient to introduce the new variable $t=(1-y)/\kappa$ in both integrals
in Eq.~(\ref{eq15}). In terms of the new variable these integrals take the form
\begin{eqnarray}
&&
Y_{00}=-\kappa^2\int_{-1}^{1}\frac{dt}{e^{\beta}e^{-\beta\kappa t}+1}
\sqrt{1-t^2},
\nonumber \\
&&
Y=-\kappa^2\int_{-1}^{1}\frac{dt}{e^{\beta}e^{-\beta\kappa t}+1}
\frac{t^2}{\sqrt{1-t^2}}.
\label{eq25}
\end{eqnarray}
\noindent
Taking into account Eqs.~(\ref{eq9}) and (\ref{eq12}),  for the parameter
$\beta\kappa$ we have
\begin{equation}
\beta\kappa=\frac{c\tilde{v}_F\hbar k}{2k_BT}\leq\tilde{v}_F
\frac{\hbar\omega}{2k_BT}.
\label{eq26}
\end{equation}
\noindent
As a result, the integrals (\ref{eq25}) can differ from zero only in the region
of the plane $(\omega,T)$ where $\beta\kappa\ll 1$. This is apparent from the fact
that, for example, for $\beta\kappa>0.1$ it occurs
\begin{equation}
0.1<\beta\kappa<\tilde{v}_F\beta
\label{eq27}
\end{equation}
\noindent
 and, thus, $\beta>30$. In this case both
functions under the integrals in Eq.~(\ref{eq25}) are of the order of exp(--30).

We arrive at the conclusion that in the region of $(\omega,T)$ plane, where the
thermal correction contributes to the result, the function under the integrals in
Eq.~(\ref{eq25}) containing exponents can be expanded in powers of the small
parameter $\beta\kappa$:
\begin{eqnarray}
&&
\frac{1}{e^{\beta}e^{-\beta\kappa t}+1}=\frac{1}{e^{\beta}+1}+
\beta\kappa\frac{e^{-\beta}}{(1+e^{-\beta})^2}t
\nonumber \\
&&~~~~~
+(\beta\kappa)^2\frac{e^{-\beta}(1-e^{-\beta})}{2(1+e^{-\beta})^3}t^2
+O[(\beta\kappa)^3].
\label{eq28}
\end{eqnarray}
\noindent
Substituting this in the first line of Eq.~(\ref{eq25}) and performing the
integration, we arrive at
\begin{equation}
Y_{00}=-\kappa^2\left[\frac{\pi}{2(e^{\beta}+1)}+
\kappa^2\frac{\pi\beta^2e^{-\beta}(1-e^{-\beta})}{16(1+e^{-\beta})^3}
+O(\kappa^4)\right].
\label{eq29}
\end{equation}
\noindent
Note that the terms in Eq.~(\ref{eq28}) containing the odd powers of the small
parameter $\beta\kappa$ (and, thus, the odd powers of $t$) do not contribute to
the result (\ref{eq29}). In a similar way, substituting Eq.~(\ref{eq28}) in
the second line of Eq.~(\ref{eq25}) and integrating, we find
\begin{equation}
Y=-\kappa^2\left[\frac{\pi}{2(e^{\beta}+1)}+
\kappa^2\frac{3\pi\beta^2e^{-\beta}(1-e^{-\beta})}{16(1+e^{-\beta})^3}
+O(\kappa^4)\right].
\label{eq30}
\end{equation}

Now we substitute Eqs.~(\ref{eq29}) and (\ref{eq30}) in Eq.~(\ref{eq24}),
expand $\eta$ in powers of a small parameter $\kappa$ defined in Eq.~(\ref{eq9})
and combine all terms with similar powers of $\kappa$. The result is
\begin{equation}
{\rm Re}\sigma_{\|(\bot)}(\omega,k,T)=
{\sigma_0}\left[1-\frac{2}{e^{\beta}+1}-\kappa^2C_{\|(\bot)}(\beta)
+O(\kappa^4)\right],
\label{eq31}
\end{equation}
\noindent
where the functions $C_{\|}$ and $C_{\bot}$ are given by
\begin{eqnarray}
&&
C_{\|}(\beta)=\frac{\beta^2e^{-\beta}(1-e^{-\beta})+
2(1+e^{-\beta})^2(1+3e^{-\beta})}{4(1+e^{-\beta})^3},
\nonumber \\
&&
C_{\bot}(\beta)=\frac{3\beta^2e^{-\beta}(1-e^{-\beta})-
2(1+e^{-\beta})^2(1+3e^{-\beta})}{4(1+e^{-\beta})^3}.
\label{eq32}
\end{eqnarray}
\noindent
According to Eq.~(\ref{eq9}), the dependence of the real parts of conductivities
(\ref{eq31}) on $k$ is contained only in the quantities $\kappa\leq\tilde{v}_F$,
whereas according to Eq.~(\ref{eq12}) the dependence on $\omega$ and $T$ is
determined  by the parameter $\beta$. Taking into account that the relative
contribution of the second order terms in Eq.~(\ref{eq31}) does not exceed $10^{-4}$,
one arrives at \cite{52b}
\begin{equation}
{\rm Re}\sigma_{\|(\bot)}(\omega,T)\approx
{\sigma_0}\left[1-\frac{2}{e^{\hbar\omega/(2k_BT)}+1}\right]=
\sigma_0\tanh\frac{\hbar\omega}{4k_BT}.
\label{eq33}
\end{equation}
\noindent
The real part of $\sigma_{\|(\bot)}$ is approximately equal to $\sigma_0$ in
the limiting case of high frequencies $\omega\gg 2\omega_T$ and goes to zero
in the limiting case of low frequencies $\omega\ll 2\omega_T$.

Note that Eq.~(\ref{eq33}) possesses a discontinuity as a function of two
variables $\omega$ and $T$ at the point (0,0):
\begin{eqnarray}
&&
\lim_{\omega\to 0}{\rm Re}\sigma_{\|(\bot)}(\omega,T\neq 0)=0,
\nonumber \\
&&
\lim_{T\to 0}{\rm Re}\sigma_{\|(\bot)}(\omega\neq 0,T)=
{\sigma_0}.
\label{eq36}
\end{eqnarray}
\noindent
We also note that the real part of the conductivity of graphene calculated using the
polarization tensor does not contain a singular term which arises in theoretical approaches
using the Drude model \cite{53a} when the relaxation parameter goes to zero. This is
because the Dirac model for pure graphene does not take into account the collisions of
quasiparticles. In terms of the Drude model this means that the relaxation parameter is
put equal to zero from the outset.

Finally, we check that the obtained results satisfy the condition (\ref{eq18}).
{}From Eq.~(\ref{eq6a}) we have
\begin{equation}
{\rm Re}\alpha_{\|(\bot)}=-2\pi \frac{k}{\omega}{\rm Im}\sigma_{\|(\bot)},
\quad
{\rm Im}\alpha_{\|(\bot)}=2\pi \frac{k}{\omega}{\rm Re}\sigma_{\|(\bot)}.
\label{eq38}
\end{equation}
\noindent
At this point we can use only the second equality in Eq.~(\ref{eq38}) leaving the
first one for Sec.~IV. Substituting Eq.~(\ref{eq31}) in Eq.~(\ref{eq38}), one finds
\begin{equation}
{\rm Im}\alpha_{\|(\bot)}=2\pi\frac{k}{\omega}
{\sigma_0}\left[1-\frac{2}{e^{\beta}+1}-\kappa^2C_{\|(\bot)}+O(\kappa^4)\right].
\label{eq39}
\end{equation}
\noindent
As a result, using Eq.~(\ref{eq9}), we obtain
\begin{equation}
{\rm Im}\alpha_{\bot}-{\rm Im}\alpha_{\|}=
2\pi{\sigma_0}\tilde{v}_F^2c^2\frac{k^3}{\omega^3}
\left[C_{\|}-C_{\bot}\right],
\label{eq40}
\end{equation}
\noindent
i.e., the condition (\ref{eq18}) is satisfied.

\subsubsection{Numerical results}

We are coming now to numerical computations of the real part of the conductivity of
graphene obtained using the polarization tensor over a wide range frequencies and
temperatures. For this purpose one can use either the exact Eqs.~(\ref{eq24}) and
(\ref{eq25}) or the approximate Eqs.~(\ref{eq31}) and (\ref{eq32}). In both cases
the computational results are essentially independent on $k$ and, thus, the local
limit $k=0$ can be taken from the very beginning.

\begin{figure}[b]
\vspace*{-8cm}
\centerline{\hspace*{2.5cm}
\includegraphics{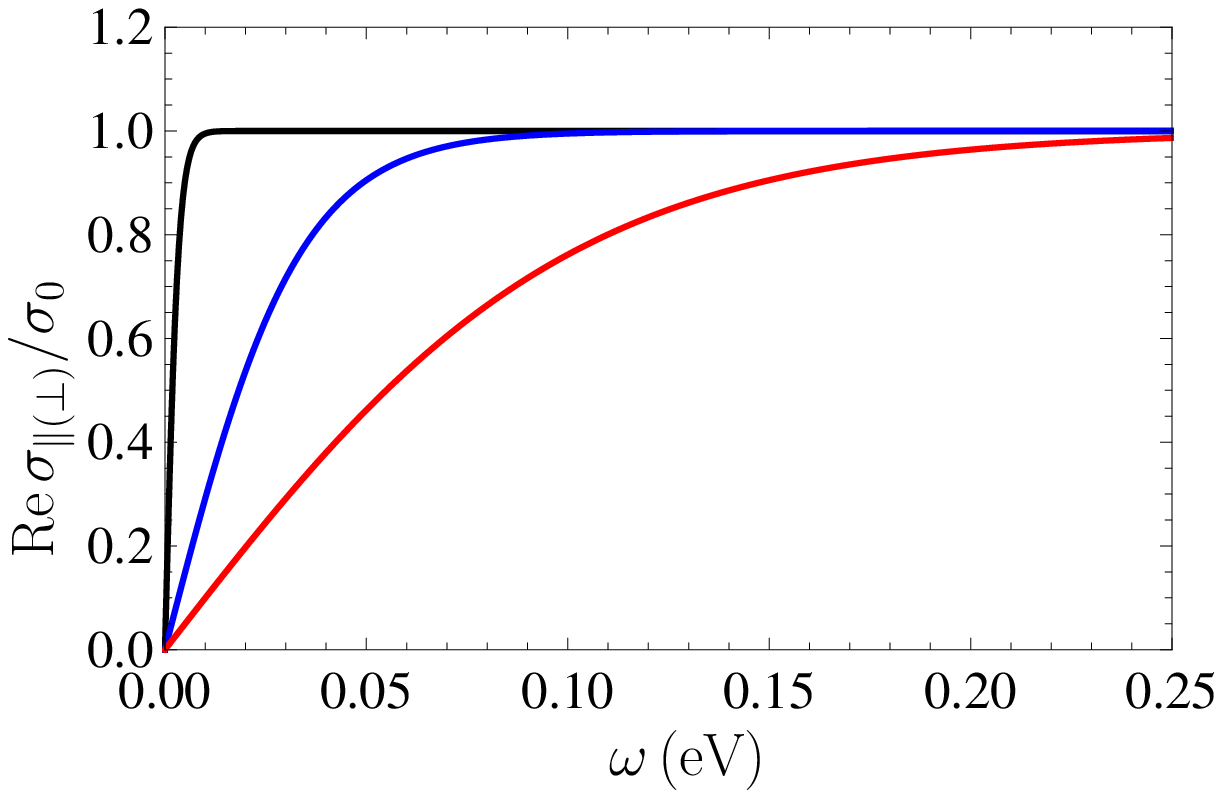}
}
\vspace*{-9cm}
\caption{\label{fg1}(Color online)
The normalized to $\sigma_0$ real part of the conductivity of pure graphene
as a function of frequency is shown by the three lines from bottom to top
computed at the temperatures $T=300\,$K, 100\,K, and 10\,K, respectively.
}
\end{figure}

In Fig.~\ref{fg1} we plot the real part of the conductivity of graphene, normalized
to the universal conductivity $\sigma_0$ defined in Eq.~(\ref{eq0}), as a function
of frequency (we remind that $1\,\mbox{eV}\approx 1.52\times 10^{15}\,$rad/s).
The lines from bottom to top are computed for the values of temperature equal to
300\,K, 100\,K and 10\,K, respectively. As is seen in this figure, with increasing
$\omega$ all lines go to a common universal limiting value $\sigma_0$ in accordance with
Eq.~(\ref{eq33}). This value of the real part of the conductivity is already achieved
at the  frequency of about 0.01\,eV when the temperature is minimal ($T=10\,$K).
With increasing temperature to 100\,K and 300\,K the conductivity $\sigma_0$ is achieved
for frequencies satisfying the inequalities $\omega>0.1\,$eV and $\omega>0.25\,$eV,
respectively. We note that at $T=10\,$K ($2k_BT\approx 0.0016\,$eV) the maximum value
of the parameter (\ref{eq12}) is
\begin{equation}
\beta_{\max}=\frac{\hbar\omega_{\max}}{2k_BT}=\frac{0.25}{0.0016}\approx 150.
\label{eq41}
\end{equation}
\noindent
Taking into account that $\kappa_{\max}=\tilde{v}_F$, one obtains
$\beta_{\max}\kappa_{\max}\approx 0.5$, i.e., in the discussed above region
$\beta\kappa>0.1$, where the thermal corrections are negligibly small in accordance
with the top line in Fig.~\ref{fg1}. With increasing $T$ the parameter $\beta$
decreases and the role of thermal effects becomes important over the wider ranges
of frequencies (see the middle and bottom lines in Fig.~\ref{fg1}).

With decreasing $\omega$ the real part of the conductivity of graphene approaches
zero in accordance with Eq.~(\ref{eq33}).
This approach becomes slower with increasing temperature.
Our computational results for the real part of the conductivity of graphene in
Fig.~\ref{fg1} are in agreement with those obtained using the tight-binding
model.\cite{57a}

\begin{figure}[b]
\vspace*{-3cm}
\centerline{\hspace*{.5cm}
\includegraphics{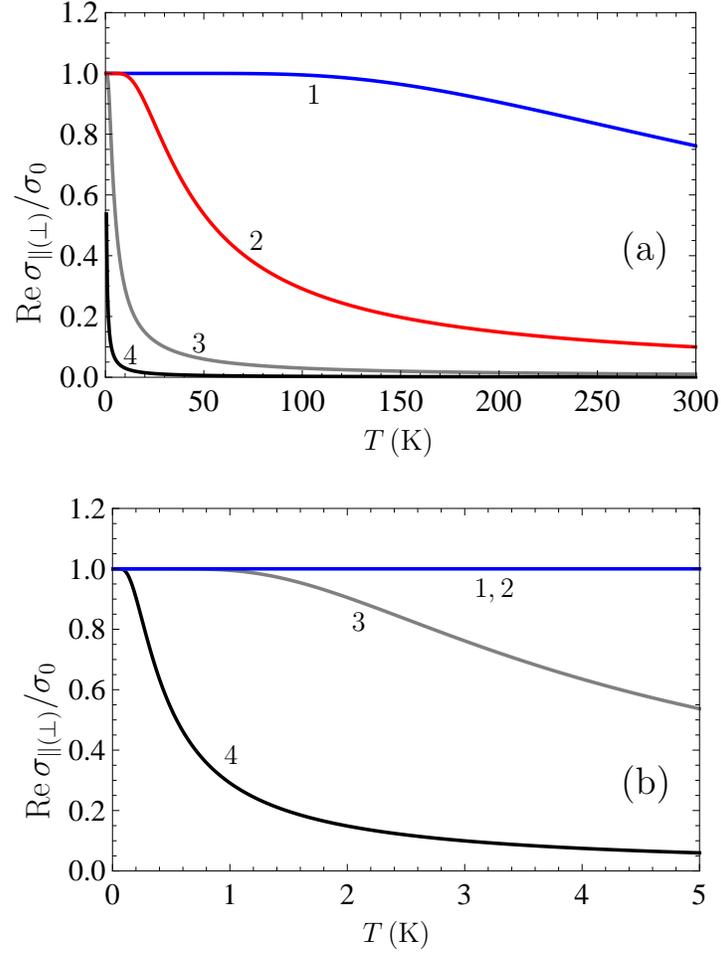}
}
\vspace*{-13cm}
\caption{\label{fg2}(Color online)
(a) The normalized to $\sigma_0$ real part of the conductivity of pure graphene
as a function of temperature in the region from 0.5\,K to 300\,K is shown by the
lines 1, 2, 3, and 4 computed at the frequencies $\omega=0.1$, 0.01, 0.001, and
0.0001\,eV, respectively. (b) The same results are shown on an enlarged scale in the
region from 0\,K to 5\,K.
}
\end{figure}

Now we compute the real part of the conductivity of graphene at fixed frequency
as a function of temperature. The computational results normalized to the universal
conductivity $\sigma_0$ are presented in Fig.~\ref{fg2}(a) in the temperature
region from 0.5\,K to room temperature. The lines 1, 2, 3, and 4 correspond to
frequencies $\omega$ equal to 0.1, 0.01, 0.001, and 0.0001\,eV, respectively.
As is seen in Fig.~\ref{fg2}(a), with increasing temperature the real part of
the graphene conductivity at all frequencies decreases and goes to the common
value equal to zero. This decrease is very slow at the largest frequency and becomes
pronounced with decreasing frequency. For better visualization of the region of
low temperatures, in Fig.~\ref{fg2}(b) we plot the same lines on an enlarged
scale in the vicinity of the absolute zero (here the lines 1 and 2 overlap).
As is seen in Fig.~\ref{fg2}(b), the real part of
the conductivity of graphene goes to $\sigma_0$
with vanishing temperature. Thus, Figs.~\ref{fg1} and \ref{fg2}(a,b) illustrate the
discontinuity of the real part of graphene conductivity (\ref{eq36}) as a function of frequency
and temperature.

\section{Imaginary part of conductivity}

Here, we consider the imaginary part of the conductivity of graphene which exists only
at nonzero temperature. In our formalism it is expressed via the real part of the
polarization tensor presented in Eqs.~(\ref{eq10})--(\ref{eq13}). We start with the
asymptotic expressions for the imaginary part of the graphene conductivity and continue
with the results of numerical computations.

\subsection{Asymptotic expressions at low and high frequencies}

Taking into account Eqs.~(\ref{eq6})--(\ref{eq8}), the imaginary parts of
the conductivities of graphene are given by
\begin{eqnarray}
&&
{\rm Im}\sigma_{\|}(\omega,k,T)=-\frac{\omega}{4\pi\hbar k^2}
{\rm Re}\Delta_T\Pi_{00}(\omega,k,T),
\nonumber \\
&&
{\rm Im}\sigma_{\bot}(\omega,k,T)=\frac{c^2}{4\pi\hbar k^2\omega}
{\rm Re}\Delta_T\Pi(\omega,k,T).
\label{eq42}
\end{eqnarray}

Substituting  Eq.~(\ref{eq10})  in  Eq.~(\ref{eq42}), one obtains
\begin{eqnarray}
&&
{\rm Im}\sigma_{\|}(\omega,k,T)=
-\sigma_0\frac{8}{\pi\kappa^2}\sum_{j=1}^{3}Z_{00}^{(j)}(\omega,k,T),
\nonumber \\
&&
{\rm Im}\sigma_{\bot}(\omega,k,T)=
\sigma_0\frac{8}{\pi\kappa^2}\sum_{j=1}^{3}Z^{(j)}(\omega,k,T),
\label{eq43}
\end{eqnarray}
\noindent
where the integrals $Z_{00}^{(j)}$ and $Z^{(j)}$ are defined in
Eqs.~(\ref{eq11})--(\ref{eq13}).

Now we consider the case of low frequencies, as compared to the thermal frequency,
i.e., $\omega\ll 2\omega_T$ ($\beta\ll 1$). In this case the asymptotic expressions
for the real part of the polarization tensor have been obtained in the literature
\begin{eqnarray}
&&
{\rm Re}\Delta_T\Pi_{00}(\omega,k,T)=-8\frac{e^2}{\hbar}\ln 2
\frac{k_BTk^2}{\omega^2}+O(\tilde{v}_F^2k^4),
\nonumber \\
&&
{\rm Re}\Delta_T\Pi(\omega,k,T)=8\frac{e^2}{\hbar}\ln 2
\frac{k_BTk^2}{c^2}+O(\tilde{v}_F^2k^4)
\label{eq44}
\end{eqnarray}
\noindent
(see Eqs.~(88) and (95) in Ref.~[\onlinecite{49}] and Eq.~(19)
in Ref.~[\onlinecite{50}]).
Note that in this case of low frequencies only the integrals $Z_{00}^{(3)}$ and
$Z^{(3)}$ in Eq.~(\ref{eq43}) give the dominant contribution, whereas the other
integrals are negligibly small. Using Eq.~(\ref{eq44}) we find the common result
for the imaginary parts of both conductivities
\begin{equation}
{\rm Im}\sigma_{\|(\bot)}(\omega,k,T)=\sigma_0\frac{8\ln 2}{\pi}
\frac{k_BT}{\hbar\omega}+O(\tilde{v}_F^2k^2).
\label{eq45}
\end{equation}
\noindent
Under the condition $\hbar\omega\lesssim 0.2k_BT$ the use of asymptotic expression
(\ref{eq45}) results in less than 1\% error in the obtained results.

We continue with the case of high frequencies satisfying the condition
 $\omega\gg 2\omega_T$ ($\beta\gg 1$). In this case the asymptotic expressions
for the real part of the polarization tensor is contained in
Eqs.~(67) and (76) in Ref.~[\onlinecite{49}]:
\begin{eqnarray}
&&
{\rm Re}\Delta_T\Pi_{00}(\omega,k,T)=48\zeta(3)e^2
\frac{k^2(k_BT)^3}{\hbar^3\omega^4}+O(\tilde{v}_F^2k^4),
\nonumber \\
&&
{\rm Re}\Delta_T\Pi(\omega,k,T)=-48\zeta(3)e^2
\frac{k^2(k_BT)^3}{c^2\hbar^3\omega^2}+O(\tilde{v}_F^2k^4),
\label{eq46}
\end{eqnarray}
where $\zeta(z)$ is the Riemann zeta function. The results (\ref{eq46}) are
determined by  the integrals $Z_{00}^{(1)}$ and $Z^{(1)}$ in Eq.~(\ref{eq43}).
Then for the imaginary parts of the conductivities one obtains
\begin{equation}
{\rm Im}\sigma_{\|(\bot)}(\omega,k,T)=-\sigma_0\frac{48\zeta(3)}{\pi}
\left(\frac{k_BT}{\hbar\omega}\right)^3+O(\tilde{v}_F^2k^2).
\label{eq47}
\end{equation}
\noindent
These expressions lead to a  less than 1\% error if
$\hbar\omega>70k_BT$.

It is possible now to check the fulfilment of the condition (\ref{eq18})
for the real parts of the polarizabilities of graphene. {}From the first
equality in Eq.~(\ref{eq38}), using Eq.~(\ref{eq45}) obtained under the
condition $\omega\ll 2\omega_T$, one finds
\begin{equation}
{\rm Re}\alpha_{\|(\bot)}=-4e^2\ln 2
\frac{k_BT}{(\hbar\omega)^2}k+O(\tilde{v}_F^2k^3).
\label{eq48}
\end{equation}
\noindent
{}From this equation we immediately arrive at
${\rm Re}(\alpha_{\bot}-\alpha_{\|})=O(\tilde{v}_F^2k^3)$
and Eq.~(\ref{eq18}) is satisfied.

In a similar way, under the condition $\omega\gg 2\omega_T$ from
Eqs.~(\ref{eq38}) and Eq.~(\ref{eq47}) one finds
\begin{equation}
{\rm Re}\alpha_{\|(\bot)}=24\zeta(3)e^2
\frac{(k_BT)^3}{(\hbar\omega)^4}k+O(\tilde{v}_F^2k^3).
\label{eq50}
\end{equation}
\noindent
{}From Eq.~(\ref{eq50}) we again conclude that
Eq.~(\ref{eq18}) is satisfied.

\subsection{Numerical results}

Numerical computations of the imaginary part of the conductivity of graphene
were performed by Eq.~(\ref{eq43}), where all notations are introduced in
Eqs.~(\ref{eq11})--(\ref{eq13}). Numerical integrations were performed at
different values of $k$ entering Eqs.~(\ref{eq11})--(\ref{eq13}) and (\ref{eq43})
only through the quantity $\kappa$. Specifically, within the intervals of
$\kappa/\tilde{v}_F$ from 1 to 0.001 and of $\beta$ from 0.0001 to 30, it was
found that at each value of $\beta$ it occurs
$\sigma_{\|}\approx\sigma_{\bot}$, and the
obtained value of $\sigma_{\|(\bot)}$ does not depend on $\kappa$ within the
relative error smaller than 0.05\%. This maximum error is achieved only in the
narrow vicinity of the value $\beta=\beta_0$ where the imaginary part of the
conductivity of graphene takes the zero value. Thus, for all practical purposes
one can consider the longitudinal and transverse conductivities of graphene
equal and compute them at any nonzero value of $k$ as a function of $\beta$
which is effectively equal to the ratio of frequency to twice temperature.

\begin{figure}[b]
\vspace*{-8cm}
\centerline{\hspace*{2.5cm}
\includegraphics{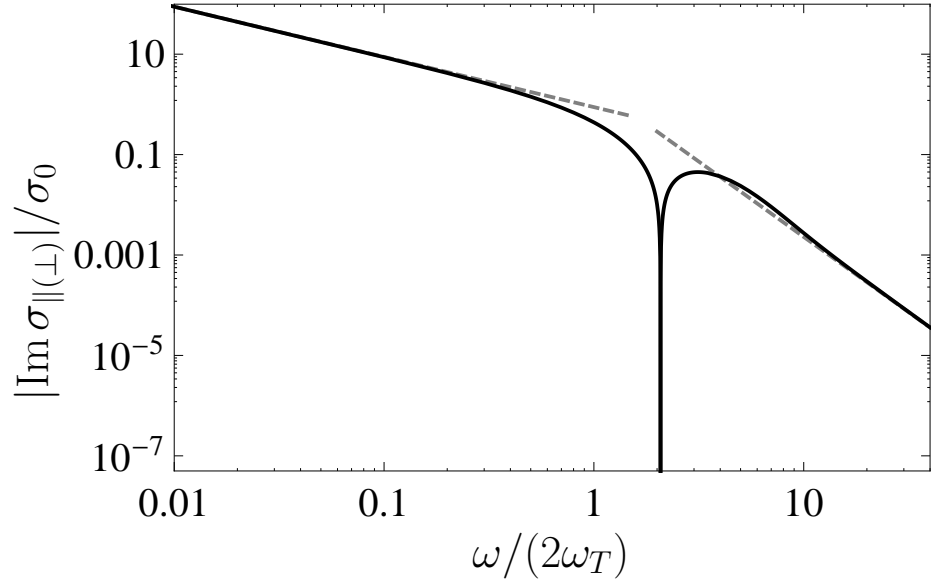}
}
\vspace*{-9cm}
\caption{\label{fg3}
The normalized to $\sigma_0$ magnitude of the imaginary
part of the conductivity of pure graphene as a function of the ratio of frequency
to twice temperature is shown by the solid line. The left and right dashed lines
present the asymptotic results at low and high frequencies, respectively.
}
\end{figure}

In Fig.~\ref{fg3} the computational results for the magnitude of the imaginary
part of graphene conductivity normalized to $\sigma_0$ are shown by the solid
line in the double logarithmic scale as a function of $\beta=\omega/(2\omega_T)$.
In the same figure the left and right dashed lines present the asymptotic results
at low and high frequencies given by Eqs.~(\ref{eq45}) and (\ref{eq47}),
respectively. As is seen in Fig.~\ref{fg3}, at both low and high frequencies the
analytic asymptotic expressions in Eqs.~(\ref{eq45}) and (\ref{eq47}) are in
a very good agreement with the results of numerical computations using the exact
formula (\ref{eq43}). It is also seen that at some ratio of frequency to
temperature $\beta_0$ the imaginary part of the conductivity of graphene takes
the zero value.

\begin{figure}[b]
\vspace*{-8cm}
\centerline{\hspace*{2.5cm}
\includegraphics{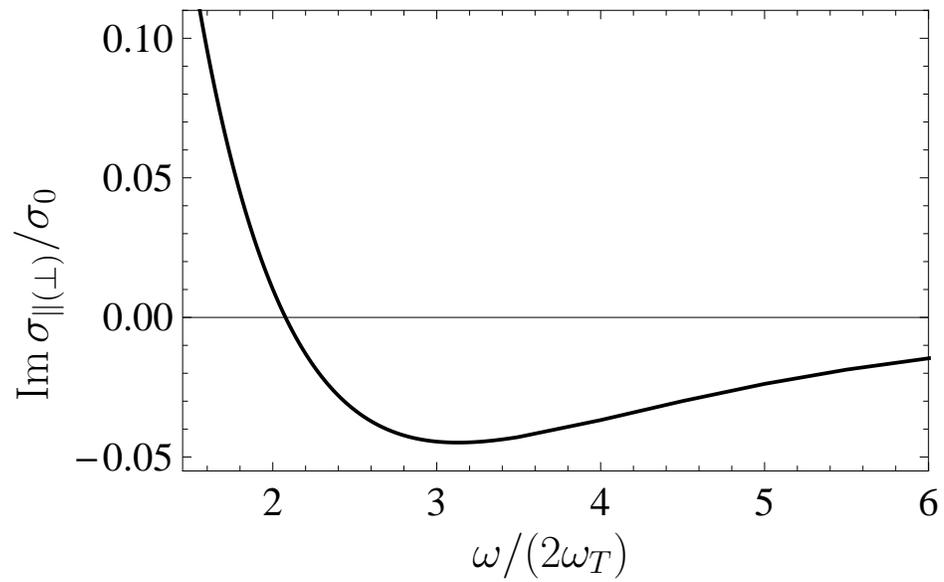}
}
\vspace*{-9cm}
\caption{\label{fg4}
The normalized to $\sigma_0$ imaginary
part of the conductivity of pure graphene as a function of the ratio of frequency
to twice temperature.
}
\end{figure}

In Fig.~\ref{fg4} we present the imaginary part of the graphene conductivity normalized
to $\sigma_0$ as a function of $\beta$ in a narrow vicinity of the point $\beta_0$,
where the imaginary part vanishes. The computations were performed for
$\kappa=0.01\tilde{v}_F$ and result in $\beta_0=2.077563$. As was noted above,
this result does not depend on the value $\kappa(k)$ to a high precision.
As can be seen in Fig.~\ref{fg4}, for $\beta<\beta_0$ the imaginary part of the
conductivity of graphene is positive and decreases from infinity to zero with
increasing $\beta$. For $\beta>\beta_0$ the imaginary part of the
conductivity is negative and approaches zero with the increase of $\beta$.
The behavior of ${\rm Im}\sigma_{\|(\bot)}$ in the limiting cases $\beta\to 0$
and $\beta\to\infty$ is in perfect agreement with the asymptotic expressions
(\ref{eq45}) and (\ref{eq47}).

\begin{figure}[b]
\vspace*{-8cm}
\centerline{\hspace*{2.5cm}
\includegraphics{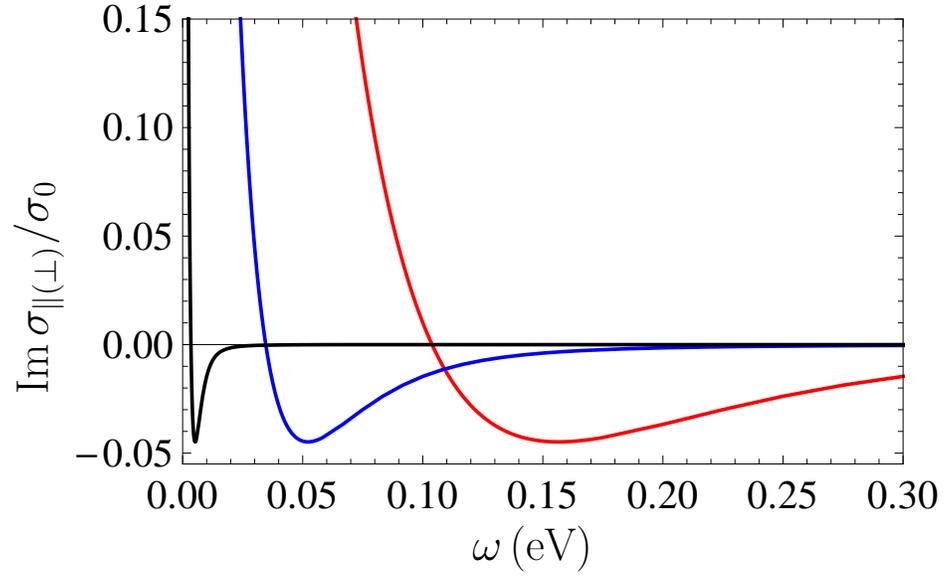}
}
\vspace*{-9cm}
\caption{\label{fg5}(Color online)
The normalized to $\sigma_0$ imaginary part of the conductivity of pure graphene
as a function of frequency is shown by the three lines from right to left
computed at the temperatures $T=300\,$K, 100\,K, and 10\,K, respectively.
}
\end{figure}

Next, we investigate the imaginary part of the conductivity of graphene at different
fixed temperatures. In Fig.~\ref{fg5} the three lines from right to left present
the computational results for ${\rm Im}\sigma_{\|(\bot)}/\sigma_0$ as the functions
of frequency at the temperatures $T=300\,$K, 100\,K, and 10\,K, respectively.
As is seen in Fig.~\ref{fg5}, with decreasing temperature from room temperature
to 10\,K, the value of frequency $\omega_0$, where the imaginary part of the conductivity
vanishes, goes to zero, and the region of frequencies, where it takes positive values,
becomes more narrow. Thus, at  $T=300\,$K, 100\,K, and 10\,K,we have
$\omega_0=0.103878$, 0.034626, and 0.0034626\,eV, respectively.

\begin{figure}[b]
\vspace*{-3cm}
\centerline{\hspace*{.5cm}
\includegraphics{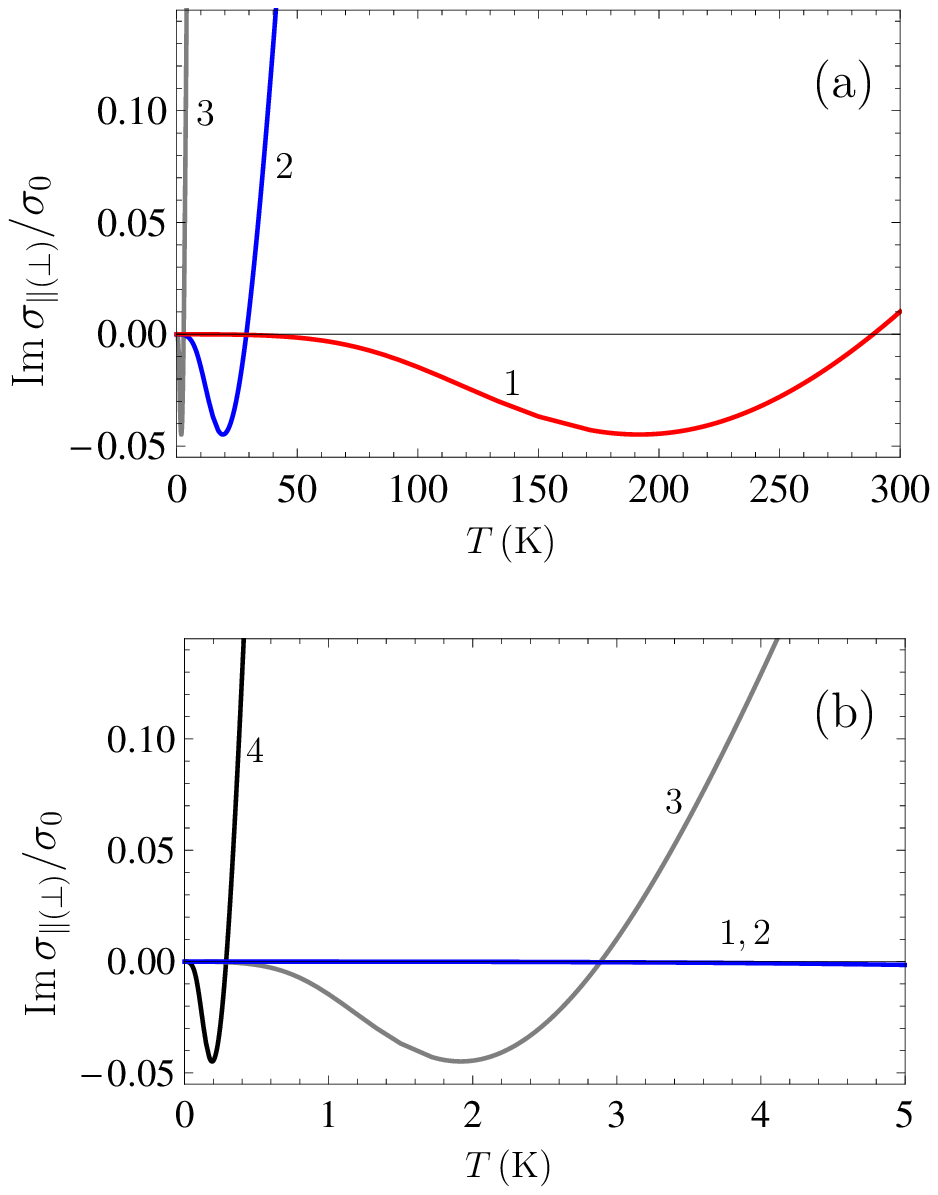}
}
\vspace*{-13cm}
\caption{\label{fg6}(Color online)
(a) The normalized to $\sigma_0$ imaginary part of the conductivity of pure graphene
as a function of temperature in the region from 0.5\,K to 300\,K is shown by the
lines 1, 2, and 3 computed at the frequencies $\omega=0.1$, 0.01, and 0.001\,eV,
 respectively. (b) The same results are shown on an enlarged scale in the
region from 0\,K to 5\,K and the line 4 is added computed at $\omega=10^{-4}\,$eV.
}
\end{figure}

Finally, we compute the imaginary part of the graphene conductivity as a function of
temperature.  In Fig.~\ref{fg6}(a) we plot ${\rm Im}\sigma_{\|(\bot)}/\sigma_0$
by the three lines numbered 1, 2, and 3 as the function of $T$ at the frequencies
$\omega=0.1$, 0.01, and 0.001\,eV, respectively.  As is seen in Fig.~\ref{fg6}(a),
at zero temperature the imaginary part of the conductivity is equal to zero, takes
negative values with increasing $T$, changes sign at some $T=T_0$ and then goes to
infinity with further increase of $T$. This is in perfect agreement with our
asymptotic expressions if to take into account that the case of low $T$ corresponds
to the asymptotic expression of high frequencies (\ref{eq47}) because at low $T$
in holds $\hbar\omega/(2k_BT)\gg 1$. In a similar way, at high  $T$
the asymptotic expression of low frequencies (\ref{eq45}) is applicable.
With decreasing $T$ the value of $T_0$ decreases linearly. Thus, it is equal to
$T_0=288,8\,$K, 28.88\,K, and 2.888\,K at $\omega=0.1$, 0.01, and 0.001\,eV, respectively.

To illustrate the behavior of the imaginary part of the conductivity at low $T$,
in Fig.~\ref{fg6}(b) we again plot the lines 1--3 already plotted in Fig.~\ref{fg6}(a),
but now over the narrow temperature interval from 0 to 5\,K and supplement them
by the line 4 plotted at $\omega=10^{-4}\,$eV. As is seen in Fig.~\ref{fg6}(b),
the lines 1 and 2 are now almost coinciding with the horizontal axis, whereas the line 4
demonstrates typical behavior of the imaginary part of the graphene conductivity
as a function of
temperature. For the line 4 we have $T_0=0.2888\,$K.

\section{Conclusions and discussion}

In the foregoing, we have investigated the conductivity of pristine (pure)
graphene in the framework of the Dirac model using the formalism of the
polarization tensor in (2+1)-dimensional space-time. Previously the
conductivity of graphene was investigated using different formalisms and
there was no complete agreement in the literature with respect to its minimal value.
According to our results, both the longitudinal and transverse conductivities
of pure graphene at zero temperature are equal to a high precision
to $\sigma_0$ defined in
Eq.~(\ref{eq0}) in agreement with a number of authors.\cite{12,18,19,21,28,53a,54}
We emphasize that the formalism of the polarization tensor is formulated
starting from the first principles of thermal quantum field theory.
It does not use any phenomenological model and
the concept of disorder with subsequent limiting transition
of the disorder parameter to zero. This in some sense simplifies calculation
of the graphene conductivity and makes it more transparent.

We have found that the real part of the conductivity of graphene at fixed
nonzero temperature goes to zero when
the frequency vanishes and achieves $\sigma_0$ with increasing frequency.
However, if the frequency is kept constant, the real part of the conductivity
of graphene goes to $\sigma_0$ and zero with decreasing and
increasing temperature, respectively.  It is shown that in the local
approximation the real parts of the longitudinal and transverse conductivities
at $T\neq 0$
are equal, and that the corrections due to a nonlocality are negligibly small
up to a very high precision. We have also obtained convenient exact
analytic expressions for the real part of the graphene conductivity. Using
these expressions, numerical computations of the real part of the
conductivity of graphene have been performed.

In this paper, the exact analytic expressions for the imaginary part of
the conductivity of graphene, which is not equal to zero at nonzero
temperature, are obtained. The asymptotic expressions for  the imaginary
part of the graphene conductivity are found at both low and high frequencies,
as compared to the thermal frequency. It is shown that the imaginary parts
of longitudinal and transverse conductivities are equal and do not depend
on the wave vector to a high precision. At some fixed temperature,
the imaginary part of the conductivity goes to infinity when the frequency
vanishes, takes the zero value at some frequency, changes its sign at higher
frequencies and goes to zero  with further increase of frequency.
At fixed frequency  the imaginary part of the conductivity takes the zero
value at zero frequency, reaches some negative minimum value with
increasing temperature and then goes to infinity when the temperature
further increases.

 The imaginary part of
the conductivity of graphene is shown to depend on the ratio of $\omega$
to $2T$. According to the obtained asymptotic expressions, at fixed $T$ it
varies from infinity at $\omega=0$ to a negative minimum value at some
fixed $\omega$, and increases to zero with further increase of $\omega$.
At fixed $\omega$ the imaginary part of the conductivity varies from zero at
$T=0$, reaches a negative minimum value at some $T$ and goes to infinity
together with $T$. Numerical computations of the imaginary
part of the conductivity are performed over the wide ranges of  frequency
and temperature.

In future it would be useful to extend the present analysis based on first
principles of thermal quantum field theory
to graphene with nonzero mass-gap parameter and chemical potential.


\newpage

\end{document}